\documentstyle[11pt,paspconf,epsf]{article}

\begin{document}
%----------------------------------------------------------------------
\title{New clues on non-standard mixing on the RGB}
{\footnote{STScI May Symposium 1998}}
\author{Corinne Charbonnel}
\affil{Laboratoire d'Astrophysique de Toulouse, CNRS UMR 5572, France}
\author{Jeffery A. Brown}
\affil{Washington State University, Pullman, USA}
\author{George Wallerstein}
\affil{Dept of Astronomy, University of Washington, Seattle, USA}

\section{Chemical anomalies in low mass red giants}
It has been known for a long time that low mass red giants 
(i.e., with initial mass lower than $\sim$ 2 M$_{\odot}$) 
present chemical anomalies which are not predicted by the standard 
evolution theory. 
In particular, the models which neglect the transport of chemicals
in the stellar radiative regions can not explain the observed behavior 
of the carbon isotopic ratio both in Pop I and II stars on the red giant 
branch (RGB), of the lithium and carbon abundances (respectively in halo and
globular cluster giants) after the completion of the first dredge-up, 
neither the evidences of O versus N anticorrelation and Na and Al versus N
correlation seen in a large number of globular cluster giants 
(see Charbonnel, Brown \& Wallerstein 1998, hereafter CBW98, for references). 
These observations suggest that low mass stars undergo a non-standard
mixing process which adds to the first dredge-up and modifies the surface
abundances.

\section{When does the extra-mixing occur?}

To throw a new light on the problem of the onset of the extra-mixing, and of
its possible metallicity dependance, we assembled accurate abundance 
observations for the $\rm ^{12}C/^{13}C$ ratio as well as for Li, C, and N 
for five field giants with [Fe/H] $\simeq - 0.6$ and two stars in the globular 
cluster 47 Tuc. 
Using the HIPPARCOS parallaxes, we could constrain the evolutionary
status of the sample stars.

Our data 
can be viewed as an evolutionary sequence, as shown in Fig.1 where 
the $\rm ^{12}C/^{13}C$ ratio is plotted against M$_{\rm bol}$ and where
the position of the luminosity function bump of 47 Tuc 
(at M$_{\rm bol}=+1.05 \pm 0.2$) is indicated.
Along this sequence, the behavior of the $\rm ^{12}C/^{13}C$ ratio is
very similar to the one observed at solar metallicity in the open
cluster M67 (Gilroy \& Brown 1991) : 
Below the bump, the observations are in agreement with standard predictions 
for dilution ($\sim 20$). Then between $\rm M_{bol}$ = +1 and +0.5, the
$\rm ^{12}C/^{13}C$ ratios drop from $\sim 20$ to near 7, i.e., well below 
the standard predicted ratio (at the same time, Li disappears and the
$\rm ^{12}C/^{14}N$ ratio diminishes by 0.2 to 0.4 dex
{\footnote{see Tables 1 and 2 in CBW98}}). 
Finaly, there is no further change in the $\rm ^{12}C/^{13}C$ ratio from
$\rm M_{bol}$ = +0.4 to $-$2.

\begin{figure}[ht]
\plotfiddle{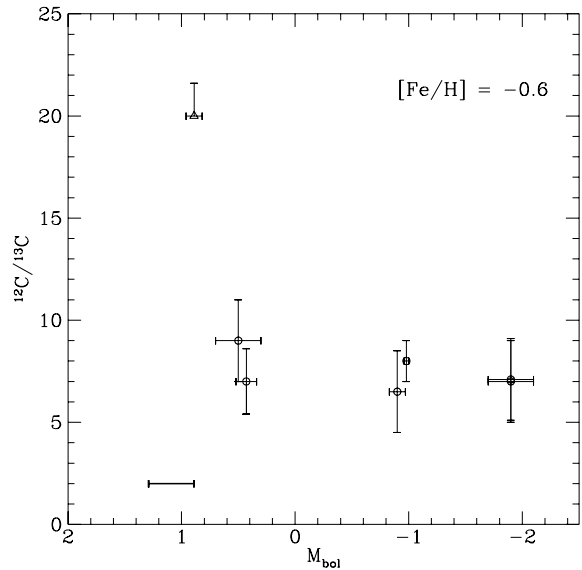}{70truemm}{0}{40}{40}{-160}{-75}
\caption{The ratio of $\rm ^{12}C/^{13}C$ is plotted against the
bolometric magnitude of our stars with moderate metal deficiency. 
The absolute magnitudes are derived from the HIPPARCOS parallaxes.
The horizontal line indicates the position of the bump in the
luminosity function of 47 Tuc (King et al. 1985).}
\end{figure}

These results show that the extra-mixing process which leads to very
low $\rm ^{12}C/^{13}C$ ratios in metal-deficient low-mass evolved stars 
becomes efficient exactly after the luminosity function bump, i.e., the 
evolutionary point where the hydrogen-burning shell crosses the chemical 
discontinuity created by the outward moving convective envelope. 
This confirms the inhibiting effect of molecular weight (or $\mu$)
barriers against the development of the extra-mixing (see Sweigart \&
Mengel 1979; Charbonnel 1994, 1995).
Since stars more massive than $\sim$ 1.7 - 2.2 M$_{\odot}$ do not reach 
the bump, they are not expected to experience this extra-mixing on the RGB
(this is confirmed by the observations in open clusters; Gilroy 1989).

\section{How many low mass stars do undergo the extra-mixing on the RGB?}
To answer this question, we collected in the literature all the
observations of the carbon isotopic ratio in field and cluster giant stars
(see Charbonnel \& Dias 1998).
For the whole sample stars, M$_{\rm V}$ values were determined thanks to
the HIPPARCOS parallaxes.
In the [Fe/H] range we consider, M$_{\rm V}^{\rm bump}$ is higher than
$\sim$ -0.2.

\begin{figure}[ht]
\plotfiddle{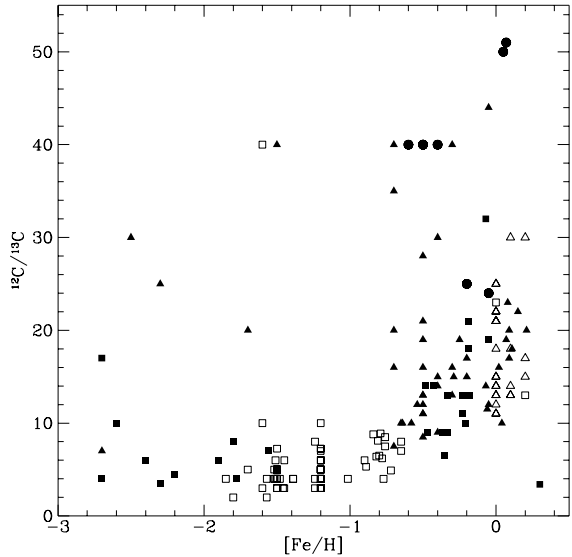}{70truemm}{0}{40}{40}{-160}{-75}
\caption{Carbon isotopic ratio collected in the literature as a function
of [Fe/H], for field and cluster giants (black and white symbols
respectively). The circles correspond to stars with M$_{\rm V}$ higher
than 2, the triangles to with M$_{\rm V}$ between 0 and 2, and the
squares to the brightest stars with M$_{\rm V}$ lower than 0}
\end{figure}

Above the bump (squares in Fig.2), i.e., the evolutionary point 
where the extra-mixing can occur, the disagreement between the post
dredge-up standard predictions ($^{12}$C/$^{13}$C around 20-30) and the 
observations appears in 96$\%$ of the low mass giants. 
This high value for the fraction of low mass stars that experience an 
extra-mixing process on the RGB and are thus expected to destroy their 
$^3$He at this evolutionary phase (Hogan 1995, Charbonnel 1995)
satisfies the galactic requirements for the evolution of the $^3$He 
abundance (see Galli et al. 1997 and Charbonnel \& Dias 1998). 

\section{Critical $\mu$-gradient}
As can be seen in Fig.2, the final $\rm ^{12}C/^{13}C$ ratios are lower
in the more metal-poor stars. The observations at various metallicities
provide precise clues on the extension of the extra-mixed region, 
and on the $\mu$-barriers that may shield the central regions of a star 
from extra-mixing. 

From evolutionary models of stars with various masses and metallicities,
we showed that, in order to account for the $^{12}$C/$^{13}$C ratios
observed in stars of different metallicities, the extra-mixing must extend 
from the base of the external convection zone down to a region where the
gradient of molecular weight is equal to $ \sim 1.5 \times 10^{-13}$. 
This value for the ``observational" critical $\mu$-gradient, ($\nabla
\ln \mu)_{c,obs}$, which appears to shield the central regions of the star 
from extra-mixing agrees with the one which is expected to stabilize 
meridional circulation. 
This result indicates that the extra-mixing on the RGB could be related to 
rotation-induced processes.
If we assume that such a $\mu$-barrier can not be penetrated, then
the energy production in the hydrogen-burning shell should not be
affected.

Let us note that, on the main sequence, $\nabla \ln \mu$ becomes higher
than ($\nabla \ln \mu)_{c,obs}$ in the very external part of the low
mass stars, and no extra-mixing is expected to occur in the stellar region 
of energy production. 
This explains the perfect agreement between standard theoretical
dilution and observations of $\rm ^{12}C/^{13}C$ for stars that have not
yet reached the bump.
This is also in agreement with the best solar models
(for what concerns helioseismological comparison and agreement with Li and 
Be observations; Richard et al. 1996) 
that include both element segregation and rotation-induced mixing which
is cut-off when the $\mu$-gradient becomes $\geq$ to 1.5 - 4 $\times 10^{-13}$.

\section{Conclusions}

The physics of the extra-mixing process in low mass stars has to be better 
understood. Detailed simulations, with a consistent treatment of the 
transport of matter and angular momentum, have to be carried out for 
different stellar masses and metallicities, and various mass loss and 
rotation histories.
The impact of this process on the behavior of various elements in RGB stars
(C $\searrow$, Na $\nearrow$, O $\searrow$, Al $\nearrow$, Mg O $\searrow$)
and on the precise yields of $^3$He has to be investigated
in details. Consequences for the energy production in the HBS and for
the HB morphology may not be neglected.


\begin{references}\small
\reference Charbonnel, C. 1994, \aap, 282, 811
\reference Charbonnel, C. 1995, \apj, 453, L41
\reference Charbonnel C., Brown J.A., Wallerstein G., 1998, A\&A 332, 204, CBW98
\reference Charbonnel C., Dias do Nascimento J., 1998, A\&A 337, in press
\reference Galli D., Stanghellini L., Tosi M., Palla F., 1997, ApJ 477, 218
\reference Gilroy, K. K. 1989, \apj, 347, 835
\reference Gilroy, K. K., Brown J. A., 1991, \apj, 371, 578
\reference Hogan C.J., 1995, ApJ Letters 441, 17
\reference King, C. R., Da Costa, G. S., and DeMarque, P. 1985 \apj, 299, 674
\reference Richard, O., Vauclair, S., Charbonnel, C., \& Dziembowski, W.A.
1996, A\&A, 312, 1000
\reference Sweigart, A. V., and Mengel, J. G. 1979, \apj, 229, 624
\end{references}
\end{document}